\documentclass[conference]{IEEEtran}
\IEEEoverridecommandlockouts
\usepackage{cite}
\usepackage{amsmath,amssymb,amsfonts}
\usepackage{algorithmic}
\usepackage{graphicx}
\usepackage{textcomp}
\usepackage{xcolor}
\usepackage{comment}

\usepackage{hyperref}
\hypersetup{
    colorlinks=true,
    linkcolor=blue,
    citecolor=blue,
    urlcolor=black
}

\def\BibTeX{{\rm B\kern-.05em{\sc i\kern-.025em b}\kern-.08em
    T\kern-.1667em\lower.7ex\hbox{E}\kern-.125emX}}
\begin{document}

\title{Severity Classification of Rotor Inter-Turn Short Circuits Using Eddy-Current Vibration Signals
}

\author{
\IEEEauthorblockN{Ondřej Rozhon,  Serge Pacome Bosson, Štěpán Janouš, Jakub Ševčík, Zdeňek Peroutka}
\IEEEauthorblockA{\textit{Faculty of Electrical Engineering} \\
\textit{University of West Bohemia} \\
Pilsen, Czech Republic \\
rozhon@fel.zcu.cz, bosson@fel.zcu.cz, sjanous@fel.zcu.cz, jsevcik@fel.zcu.cz, pero@fel.zcu.cz}
}
\maketitle

\begin{abstract}

Rotor inter-turn short-circuit (ITSC) faults in synchronous generators introduce electromagnetic asymmetries that can lead to torque ripple, unbalanced magnetic pull, and progressive mechanical degradation. While most existing studies focus on binary classification and severe fault conditions, the assessment of incipient rotor ITSC severity using displacement-sensitive vibration measurements remains relatively underexplored.

This paper proposes a vibration-based diagnostic framework for multi-class severity classification of rotor ITSC by integrating an eddy-current displacement sensor with physically motivated feature extraction. An 18-dimensional hybrid feature set is designed to characterize electromechanical modulations induced by rotor electromagnetic asymmetry.

Using an XGBoost classifier with leave-one-out cross-validation, the proposed approach achieved 90.56~\% overall accuracy, including 99~\% recall for healthy operation and 87~\% recall for mild fault conditions. The results suggest that displacement-sensitive vibration analysis enables effective severity-aware diagnosis of rotor ITSC with minimal sensor requirements.
\color{black}
\end{abstract}


\section{Introduction}

Rotor inter-turn short-circuit (ITSC) faults are among the most critical and difficult-to-detect electrical faults in synchronous generators. Such faults introduce rotor winding asymmetry, leading to unbalanced magnetic pull, torque ripple, and increased mechanical stress that can progressively degrade generator performance and reliability \cite{Klempner, Tavner}. If left undetected, rotor ITSC faults may propagate and result in severe secondary damage or forced outages, making early diagnosis essential for secure and economical power system operation~\cite{NandiReview2005}.

A wide range of diagnostic techniques has been proposed for rotor ITSC detection. Conventional approaches rely on detection coils or auxiliary windings to monitor variations in the rotor magnetic field. Although effective under severe fault conditions, these methods are sensitive to operating point variations, require intrusive installation, and often depend on manually defined thresholds that limit robustness and automation \cite{Albright}. Alternative techniques based on virtual power estimation \cite{VirtualPower}, shaft voltage monitoring \cite{ShaftVoltage}, or negative-sequence current analysis have also been reported; however, these approaches typically require specialized sensors or complex signal processing, restricting their practical deployment in industrial environments.

To overcome the limitations of purely electrical monitoring, vibration-based diagnostics have been increasingly investigated in context with rotor fault detection in synchronous generators. Rotor inter-turn short-circuit (ITSC) faults introduce electromagnetic asymmetry that produces torque ripple and unbalanced magnetic pull (UMP), which can excite measurable rotor vibration responses \cite{HeEnergies2023,YuanMPE2021}. Recent studies have further explored the combination of vibration with electrical variables to enhance sensitivity to early defects. For example, \cite{FangEnergyReports2023}~proposes a vibration-current fusion approach based on excitation-current prediction and vibration correlation, validated using field data from a large turbogenerator. Nevertheless, most vibration-based investigations rely on conventional accelerometer measurements and emphasize fault detection through RMS values or spectral indicators, rather than providing severity-aware assessment \cite{dosSantosCBMAG2022}. As a result, the existing literature offers limited support for maintenance strategies that require discrimination and tracking of incipient rotor ITSC progression.

For condition-based maintenance applications, binary fault detection is often insufficient. Maintenance decisions depend critically on fault severity: incipient rotor ITSC may be monitored until a scheduled outage, whereas advanced faults require immediate intervention. Despite this practical need, severity-aware or multi-class classification of rotor ITSC remains insufficiently explored in the literature, particularly when relying on vibration-based measurements. In addition, conventional vibration feature extraction frequently depends on predetermined narrow frequency bands and amplitude-based metrics (e.g., RMS or peak values), which may become insensitive to early-stage faults due to mechanical damping and measurement noise.

Rotor ITSC faults primarily induce low-frequency electromechanical effects associated with unbalanced magnetic pull and torque ripple. Thus, eddy-current displacement sensors are well suited for capturing such low-frequency motion and provide a non-intrusive measurement solution that is compatible with industrial synchronous generator monitoring environments.

This paper proposes a machine learning-based diagnostic framework for multi-class severity classification of rotor ITSC in synchronous generators. By integrating eddy-current displacement vibration measurements with hybrid spectral, temporal complexity, and statistical feature extraction, the proposed approach aims to capture subtle electromechanical modulations associated with different fault severity levels. A supervised learning model is employed to map these features to discrete severity classes, enabling interpretable and practical fault assessment using minimal sensor infrastructure.

The main contributions of this paper are summarized as follows:
\begin{itemize}
    \item A multi-class diagnostic framework for rotor inter-turn short-circuit faults, enabling explicit severity discrimination beyond conventional binary detection.
    \item Application of eddy-current displacement vibration sensing for rotor ITSC diagnosis, targeting low-frequency electromechanical effects induced by rotor electromagnetic asymmetry.
    \item A physically motivated hybrid feature extraction strategy combining optimized spectral integration, temporal complexity measures, and statistical morphology features.
    \item Experimental validation on a laboratory synchronous generator demonstrating effective severity-aware diagnosis using minimal sensor infrastructure.
\end{itemize}

\color{black}
\section{Proposed Diagnostic Framework}

The proposed diagnostic framework combines complementary information from both the frequency and time domains in order to capture subtle fault-induced signatures that cannot be reliably detected using conventional single-domain approaches. By integrating optimized spectral analysis with temporal complexity features and advanced machine learning, the proposed methodology aims to achieve robust fault classification under varying operating conditions. The overall diagnostic procedure consists of four main stages: (i) signal conditioning and preprocessing, (ii) optimized spectral feature extraction, (iii) temporal and statistical feature extraction, and (iv) supervised classification using an XGBoost model.

Each stage is designed to enhance sensitivity to ITSC-related phenomena while maintaining robustness against noise, load variations, and speed fluctuations.

Raw vibration signals acquired from the machine are first preprocessed to remove measurement noise and high-frequency disturbances that do not carry diagnostic information. This is achieved using a Savitzky–Golay smoothing filter, which preserves important waveform characteristics such as peaks and local trends while reducing random fluctuations. The filtered signal $x_{\text{filt}}[n]$ is obtained as
\begin{equation*}
x_{\text{filt}}[n] = \sum_{j=-M}^{M} c_j x_{\text{raw}}[n+j],
\end{equation*}
where $c_j$ denotes the convolution coefficients. The window length of 51 samples, i.e., $M=25$, and a third-order polynomial were adopted based on empirical tuning. This configuration provides an effective trade-off between noise suppression and signal fidelity, ensuring that fault-related modulation components are preserved for subsequent analysis.

Accurate estimation of the fundamental rotation frequency~$f_r$ is essential for reliable extraction of order-domain features. In practical measurements, the amplitude of harmonic components can vary significantly with load and fault severity, sometimes causing higher harmonics to dominate over the fundamental component. To avoid misidentification, the rotation frequency is estimated in the time domain using the autocorrelation function:

\begin{equation*}
R_{xx}[\tau] = \sum\nolimits_{n} x_{\text{filt}}[n] \cdot x_{\text{filt}}[n+\tau].
\end{equation*}

The dominant periodicity detected from $R_{xx}[\tau]$ provides a robust estimate of $f_r$ that is less sensitive to spectral distortions or amplitude fluctuations. This approach ensures consistent alignment of spectral features across different operating points and fault conditions.

\begin{figure}[!t]
    \centering
    \includegraphics[width=0.85\linewidth]{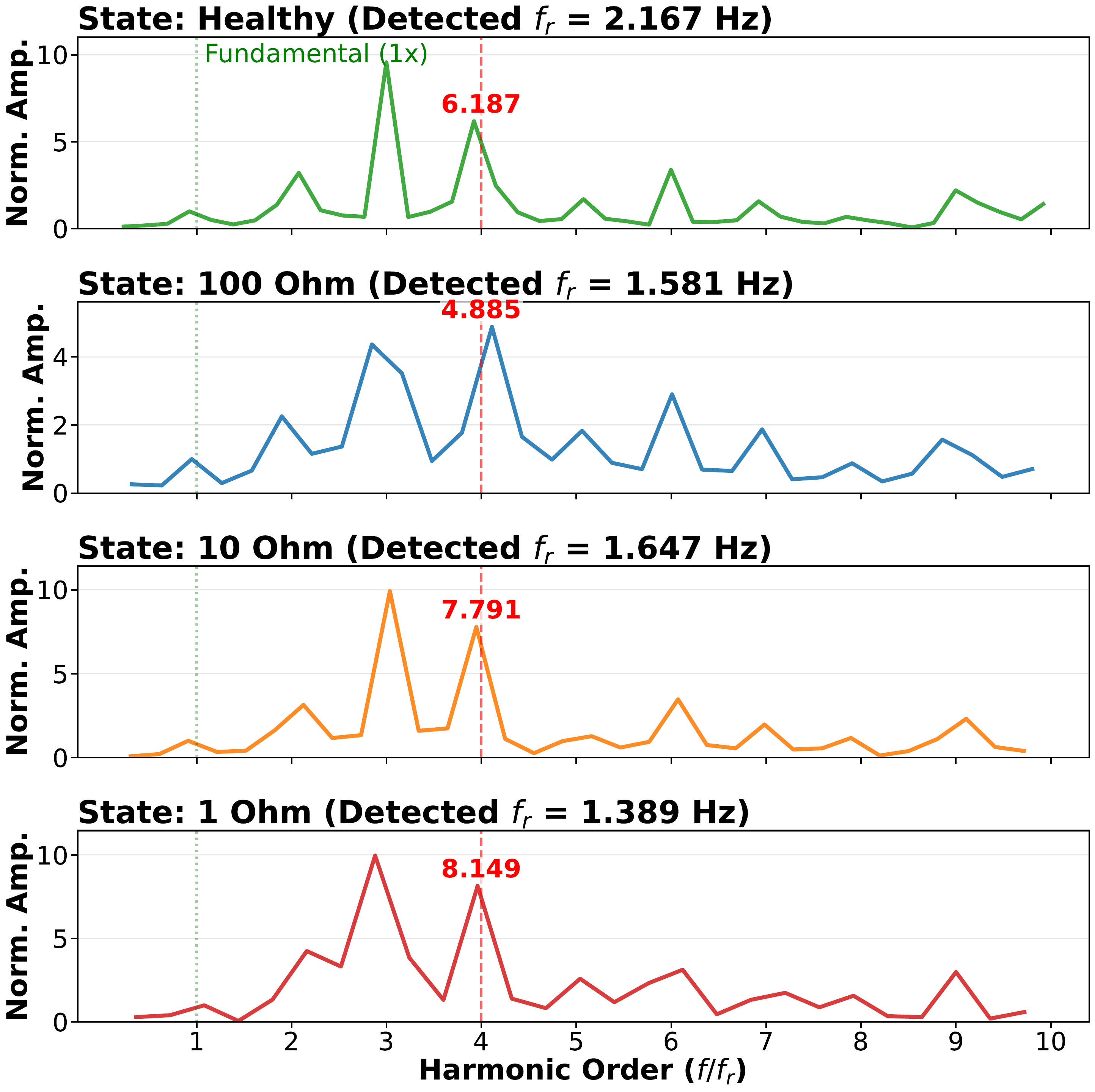}
    \caption{Order analysis of representative measurements showing normalized amplitude versus harmonic order. The 3×$f_r$ component dominates due to electromagnetic forcing, while 4×$f_r$ is selected as the primary diagnostic frequency for ITSC detection.}
    \label{fig:order_analysis2}
\end{figure}

Order-domain analysis is employed to reveal fault-related patterns that are synchronized with the mechanical rotation of the machine. As illustrated in Fig.~\ref{fig:order_analysis2}, ITSC faults do not primarily manifest as isolated spectral peaks, but rather as modulation effects that influence multiple higher-order harmonics. In particular, harmonics such as 4×, 6×, and 8× $f_r$ exhibit progressive sideband structures as fault severity increases. These sidebands are caused by periodic variations in the unbalanced magnetic pull (UMP) force generated by shorted turns. 

Although the 3×$f_r$ component is typically dominant due to electromagnetic forcing, it is less sensitive to ITSC progression. For this reason, the 4×$f_r$ harmonic is selected as the primary carrier frequency for diagnostic feature extraction, as it provides a better balance between sensitivity and stability.

\begin{figure}[!t]
    \centering
    \includegraphics[width=0.92\linewidth]{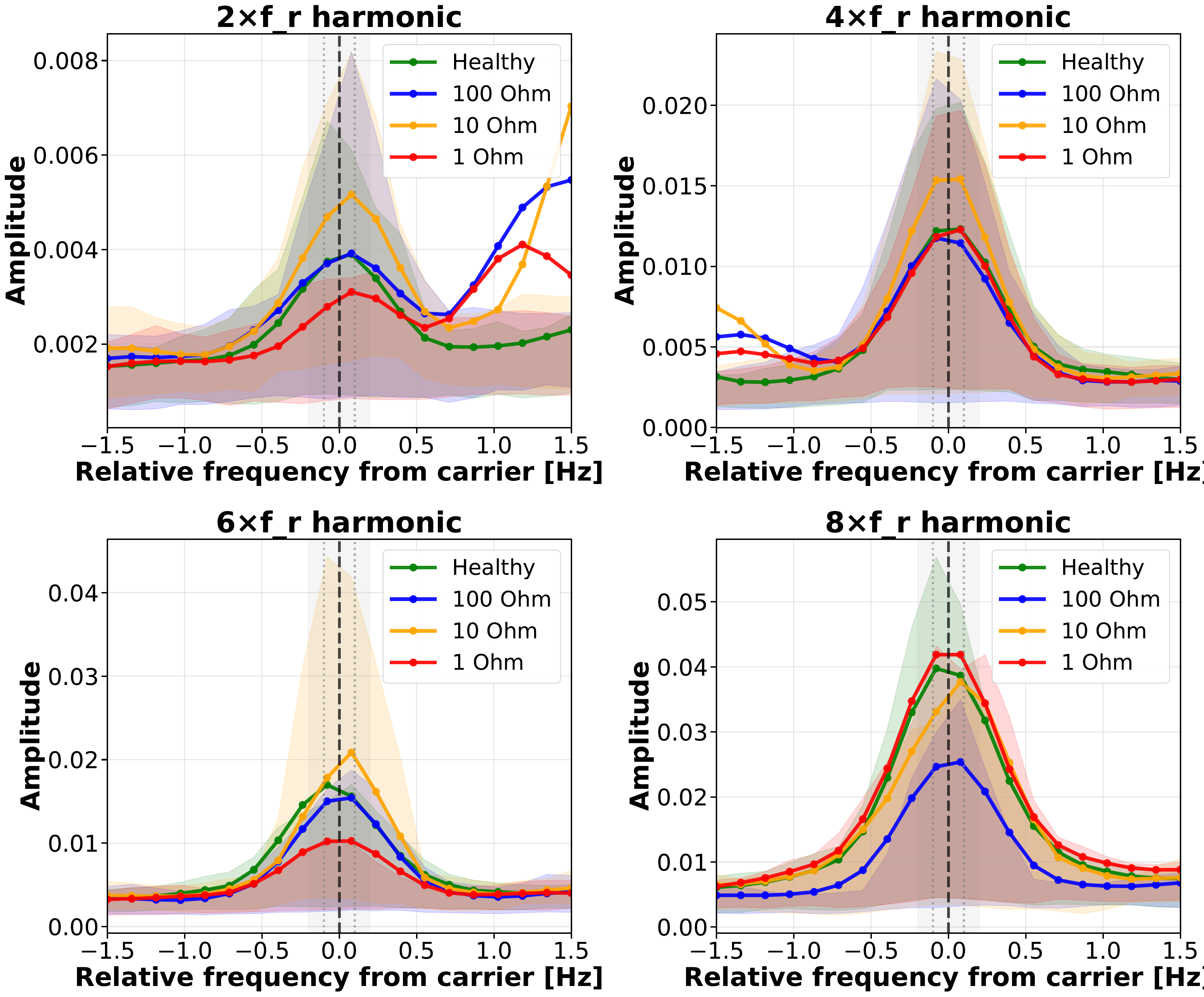}
    \caption{Ensemble-averaged spectra relative to carrier frequency for mechanical harmonics 2×, 4×, 6×, and 8× $f_r$ across all fault states. Each curve represents the mean of $n$ measurements (legend) with interquartile range (shaded regions).}
    \label{fig:averaged_spectra2}
\end{figure}

Ensemble-averaged spectra computed over all operating conditions (Fig.~\ref{fig:averaged_spectra2}) further confirm that ITSC faults induce systematic spectral evolution rather than random variations. This observation motivates the use of quantitative spectral descriptors instead of simple peak amplitude tracking.

To capture these spectral changes in a structured manner, two complementary feature types are extracted for each selected harmonic order $h \in {2, 4, 6, 8}$:

\begin{enumerate}
    \item Carrier RMS Energy:
\begin{equation*}
E_{\text{RMS}}^{(h)} = \sqrt{\frac{1}{N_w} \sum_{f \in [h f_r - 1.0, h f_r + 1.0]} |X(f)|^2},
\end{equation*}
which quantifies the overall vibration energy around the harmonic carrier frequency.

\item Sideband Ratio:
\begin{equation*}
R_{\text{SB}}^{(h)} = \frac{E_{\text{SB}}^{(h)}}{E_{\text{RMS}}^{(h)}},
\end{equation*}
where $E_{\text{SB}}^{(h)}$ represents the combined energy of the left and right modulation sidebands located at $h f_r \pm 0.1$\,Hz. This ratio captures the relative strength of fault-induced amplitude modulation independently of the overall vibration level.
\end{enumerate}

\begin{figure}[!t]
    \centering
    \includegraphics[width=0.9\linewidth]{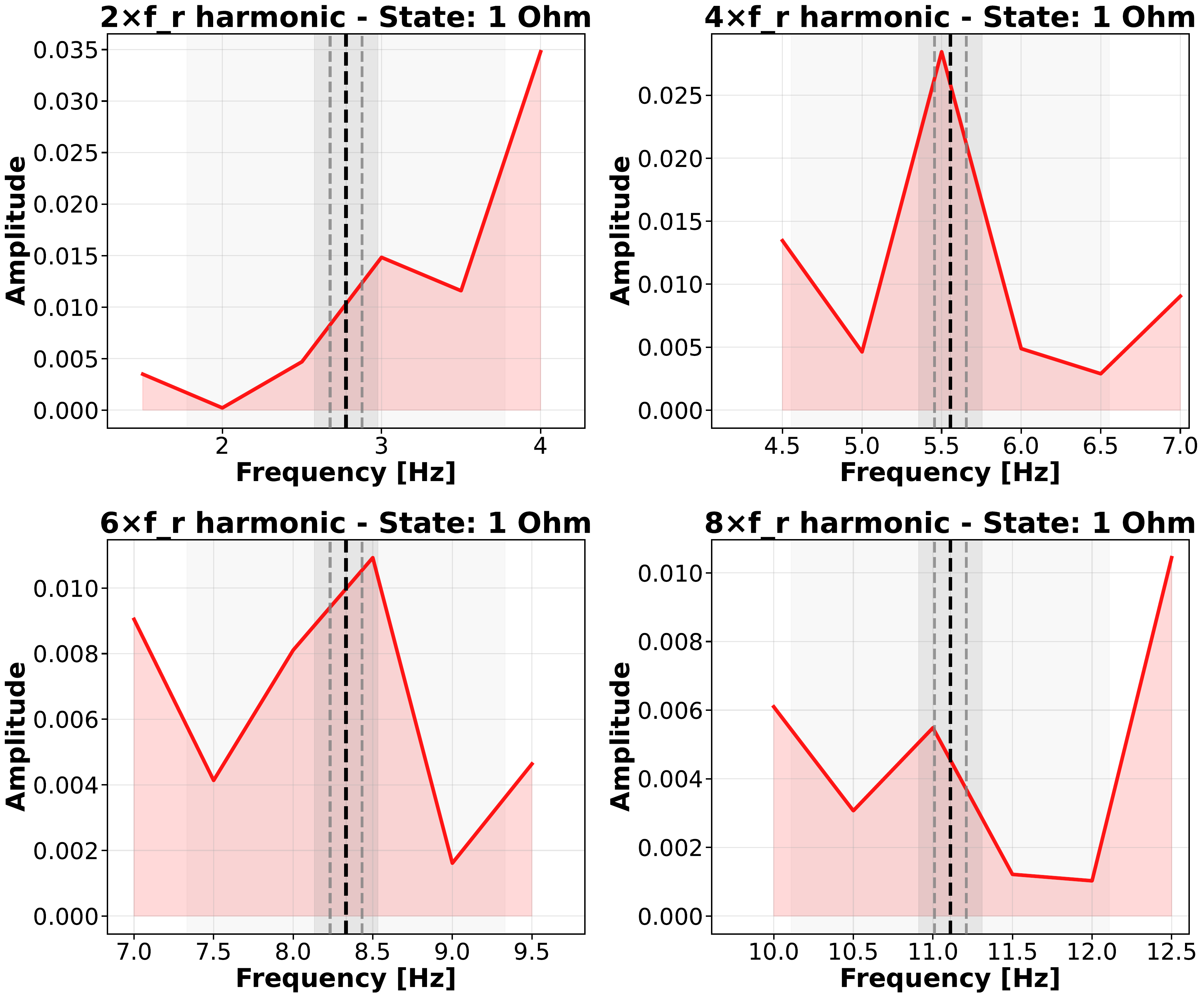}
    \caption{Feature extraction windows illustrated on a representative severe ITSC spectrum (1 Ohm state). Gray shaded regions indicate RMS integration windows (±1.0 Hz from carrier), while dashed vertical lines mark sideband detection windows (±0.1 Hz width) centered at carrier ± 0.1 Hz. The black dashed line shows the carrier frequency position at $h f_r$.}
    \label{fig:feature_windows2}
\end{figure}

The extraction strategy is illustrated in Fig.~\ref{fig:feature_windows2}. Wide RMS integration windows (±1.0\,Hz) ensure robustness to small errors in frequency estimation, while narrow sideband windows (±0.1\,Hz) focus specifically on rotation-synchronous modulation components. This procedure yields a total of eight spectral features (two per harmonic), providing a compact yet informative representation of the frequency-domain fault signatures.

While spectral features effectively describe steady-state periodic behavior, ITSC faults also influence the time-domain structure of vibration signals. To capture these effects, seven additional temporal and statistical features are extracted. These features characterize nonlinear complexity, waveform morphology, and overall signal amplitude, complementing the spectral descriptors.

\vspace{0.2cm}
\noindent\textbf{Temporal Complexity Measures:}\nobreak\\
\indent Hjorth Complexity \cite{Hjorth1970} evaluates how far a signal deviates from pure sinusoidal behavior:
\begin{equation*}
C = \sqrt{\frac{\text{Var}(x'')\text{Var}(x)}{\text{Var}(x')^2}},
\end{equation*}
where $\mathrm{Var}(\cdot)$ denotes the variance of the signal. Healthy machines exhibit relatively complex vibration patterns due to the interaction of multiple mechanical modes. As ITSC develops, the vibration becomes increasingly dominated by a periodic UMP forcing component, leading to a measurable reduction in Hjorth complexity. 

Permutation Entropy \cite{Bandt2002,Rajabi2022} measures the temporal unpredictability of the signal:
\begin{equation*}
H = -\sum\nolimits_\pi p(\pi) \ln p(\pi),
\end{equation*}
where $\pi$ denotes an ordinal pattern of length $D$, and $p(\pi)$ represents its relative probability of occurrence in the embedded time series. In this study, an embedding dimension of $D=3$ and delay $\tau=1$ were used. Fault-induced modulation may increase the irregularity of the temporal sequence, resulting in higher entropy values.

\vspace{0.2cm}
\noindent\textbf{Statistical Morphology Features:}\nobreak\\
\indent Classical statistical indicators including the root-mean-square (RMS), crest factor, peak-to-peak value, skewness, and kurtosis are also included to capture global changes in vibration behavior:
\begin{equation*}
\text{RMS}_{\text{total}} = \sqrt{\frac{1}{N}\sum\nolimits_n^N x[n]^2}, \quad\quad
\text{Crest} = \frac{\max|x|}{\text{RMS}_\text{total}},
\end{equation*}
\begin{equation*}
\text{Peak-to-Peak} = \max(x) - \min(x), \quad
\end{equation*}
\begin{equation*}
\text{Skewness} = \frac{\sum\nolimits_n^N (x[n]-\mu)^3}{N\sigma^3},~~
\text{Kurtosis} = \frac{\sum\nolimits_n^N(x[n]-\mu)^4}{N\sigma^4},
\end{equation*}
where $\mu$ denotes the mean value of the vibration signal $x$ and $\sigma$ its standard deviation. 
These metrics reflect changes in overall vibration intensity, impulsiveness, asymmetry, and distribution shape that accompany magnetic and mechanical distortions caused by ITSC.

To help the classifier distinguish fault effects from normal operational variations, two current-based features are incorporated:
\begin{equation*}
I_{\text{mean}} = \frac{1}{N}\sum\nolimits_n^N i[n], \quad
I_{\text{std}} = \sqrt{\frac{1}{N}\sum\nolimits_n^N (i[n] - I_{\text{mean}})^2}.
\end{equation*}

The mean current reflects the excitation level of the machine, while its standard deviation captures load fluctuations. Including these parameters enables the model to adapt to different operating regimes rather than learning spurious correlations.

Finally, the estimated rotation frequency $f_r$ is also included as an explicit feature, allowing the diagnostic model to account for speed-dependent variations in fault manifestation.

\begin{figure}[!t]
    \centering
    \includegraphics[width=0.78\linewidth]{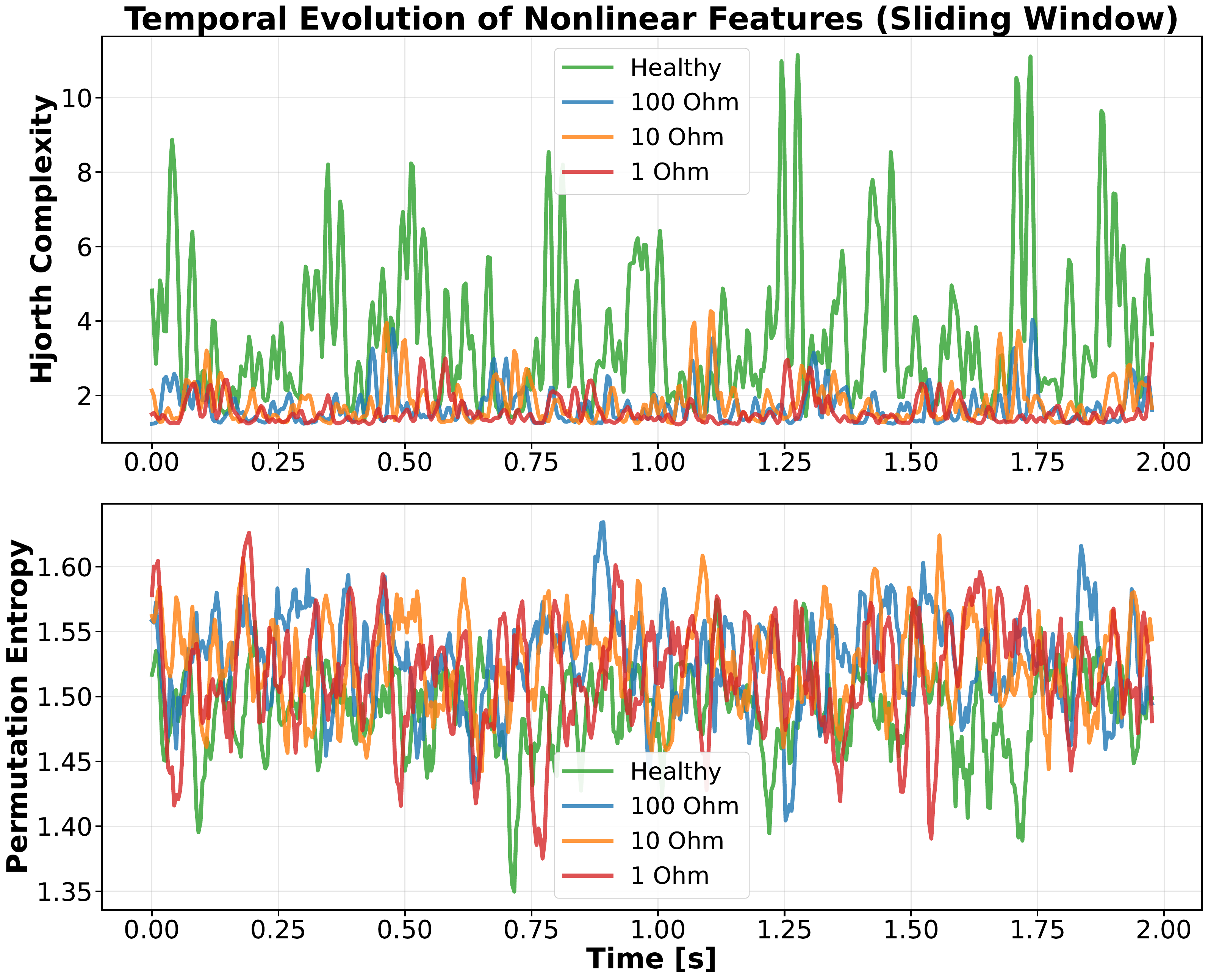}
    \caption{Temporal evolution of Hjorth Complexity and Permutation Entropy via sliding window analysis for representative measurements across fault conditions.}
    \label{fig:temporal_evolution2}
\end{figure}

Temporal behavior of key nonlinear features is analyzed using a sliding-window approach with 0.5-s windows and 0.1-s overlap. Representative evolutions of Hjorth Complexity and Permutation Entropy are shown in Fig.~\ref{fig:temporal_evolution2}, illustrating their sensitivity to fault severity.
\section{Experimental Results and Analysis}
The proposed diagnostic framework was validated on a laboratory-scale test rig designed to emulate rotor inter-turn short circuits under controlled conditions.

\subsection{Test Rig Description}
\begin{figure}[!t]
    \centering
    \includegraphics[width=0.62\linewidth]{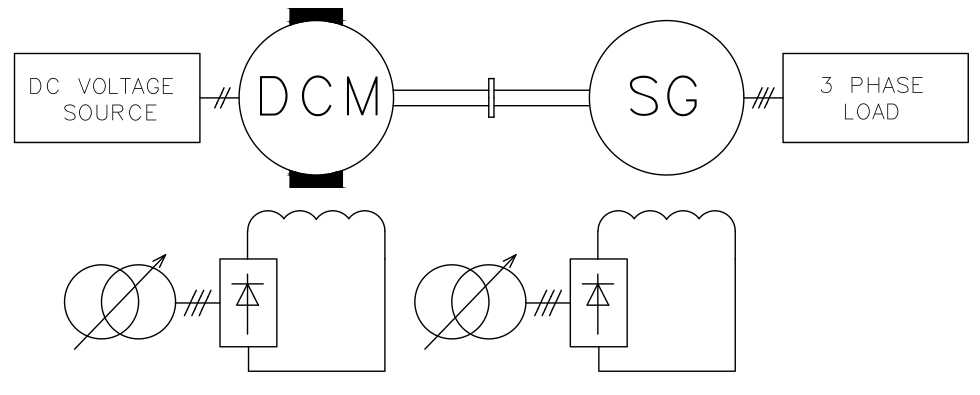}
    \caption{Schematic diagram of the laboratory test rig showing the synchronous generator, DC drive motor, excitation system, and instrumentation layout.}
    \label{fig:schematic2}
\end{figure}

The experimental setup consists of a salient-pole synchronous generator with a four-pole rotor. The generator is mechanically coupled to a DC drive motor, which serves as a controllable loading machine. The synchronous machine is connected to an isolated network supplying a resistive load bank. A schematic of the laboratory circuit is shown in Fig.~\ref{fig:schematic2}.

\begin{figure}[!t]
    \centering
    \includegraphics[width=0.7\linewidth]{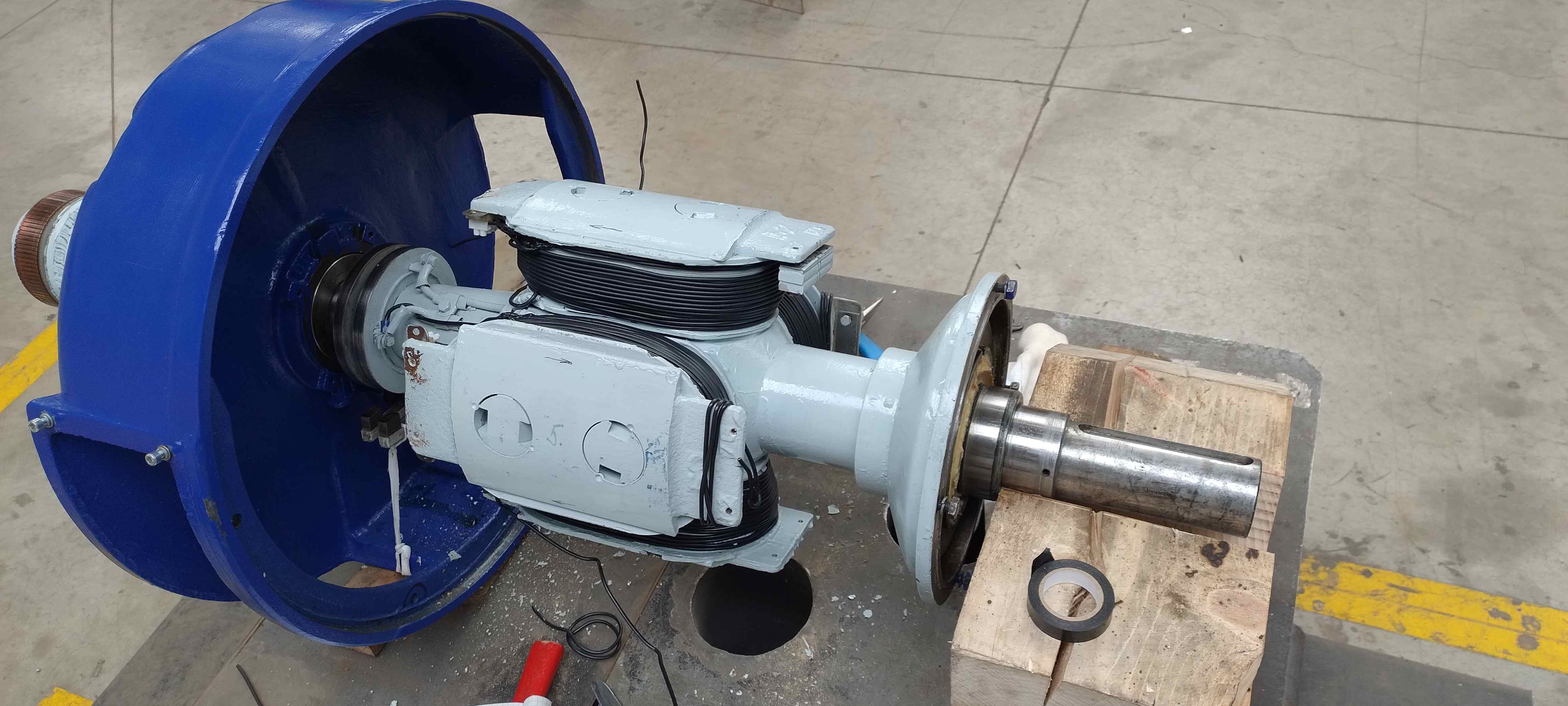}
    \caption{Modification of the selected generator's field winding to allow external taps for controlled fault injection.}
    \label{fig:BG2}
\end{figure}

To emulate realistic ITSC faults without permanently damaging the machine, a modification was made to one of the rotor poles. Taps were introduced into the field winding, enabling a specific number of turns to be by-passed externally via slip rings, as shown in Fig.~\ref{fig:BG2}.

\begin{figure}[!t]
    \centering
    \includegraphics[width=0.7\linewidth, angle=180]{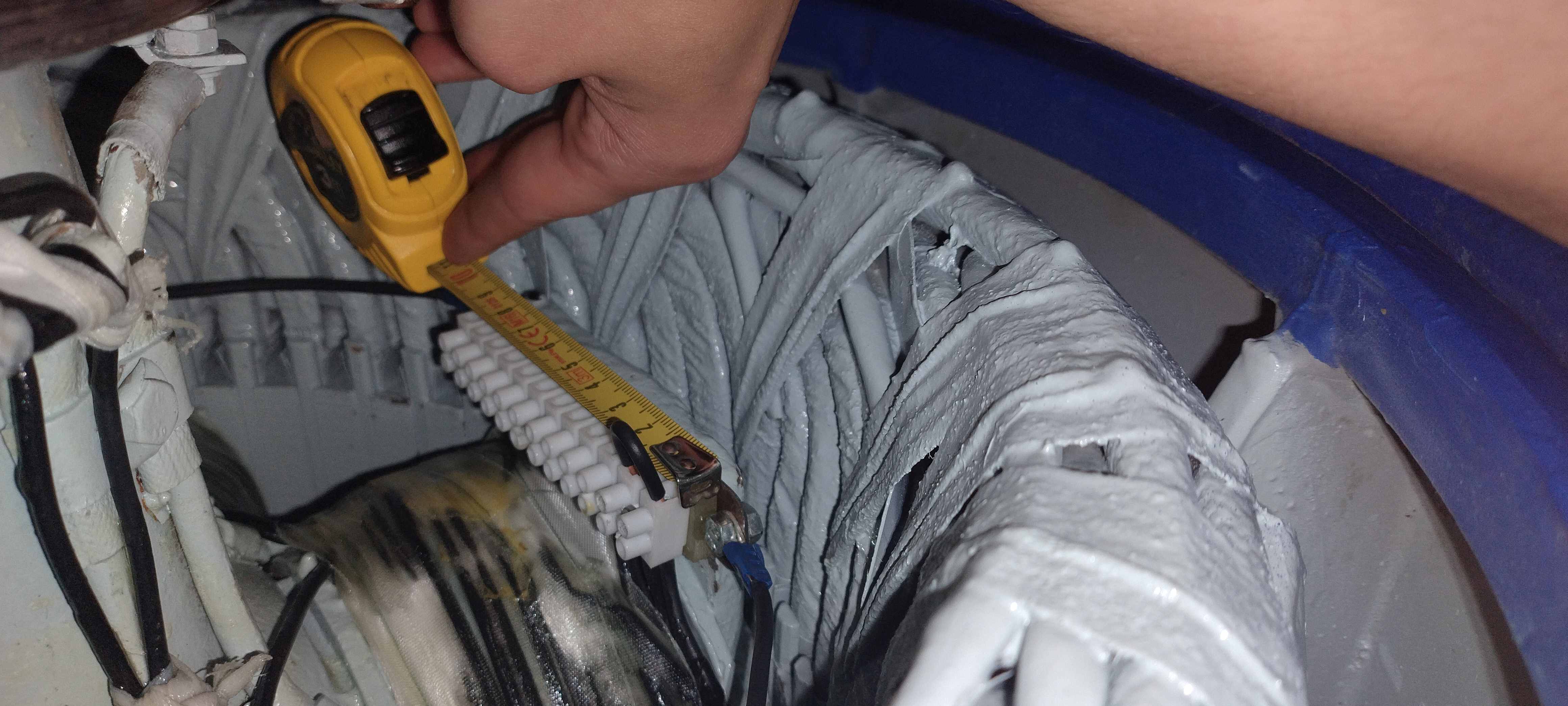}
    \caption{Auxiliary terminal block with shorting cable used for fault injection via external resistors.}
    \label{fig:svorkovnice}
\end{figure}

Fault severity is controlled by connecting external resistors to the shorting path. Three distinct fault levels were defined (i)~{\it Severe Fault} (1\,$\Omega$), which represents a near-complete insulation failure, (ii)~{\it Moderate Fault} (10\,$\Omega$), and (iii)~{\it Mild Fault} (100\,$\Omega$), which result in low circulating current and represents a case which is usually difficult to detect.
A {\it Healthy} baseline condition was also recorded (no shorted turns).

\subsection{Instrumentation and Data Acquisition}
Vibration data were acquired using a single non-contact eddy-current displacement sensor mounted radially near the rotor shaft. Unlike accelerometers, displacement sensors directly measure the shaft's relative motion, making them highly sensitive to the low-frequency mechanical excursions caused by unbalanced magnetic pull.
A current probe was used to measure the excitation current.

The dataset comprises 360 steady-state measurements acquired across a wide range of operating conditions to ensure model robustness and prevent overfitting. The experiments were conducted at various rotational frequencies and excitation current levels to generate diverse training datasets, ensuring the classifier robustness and compatibility with realistic industrial operating conditions.
The distribution of the 360 measurements across fault conditions is summarized in Table~\ref{tab:dataset2}.
\begin{table}[htbp]
\centering
\caption{Dataset Composition}
\label{tab:dataset2}
\begin{tabular}{lcc}
\hline
\textbf{Condition} & \textbf{Samples} & \textbf{Percentage} \\
\hline
Healthy           & 79  & 21.9\% \\
Mild (100~$\Omega$)   & 99  & 27.5\% \\
Moderate (10~$\Omega$) & 87  & 24.2\% \\
Severe (1~$\Omega$)    & 95  & 26.4\% \\
\hline
\textbf{Total}    & \textbf{360} & \textbf{100\%} \\
\hline
\end{tabular}
\end{table}

\begin{figure}[!t]
    \centering
    \includegraphics[width=0.8\linewidth]{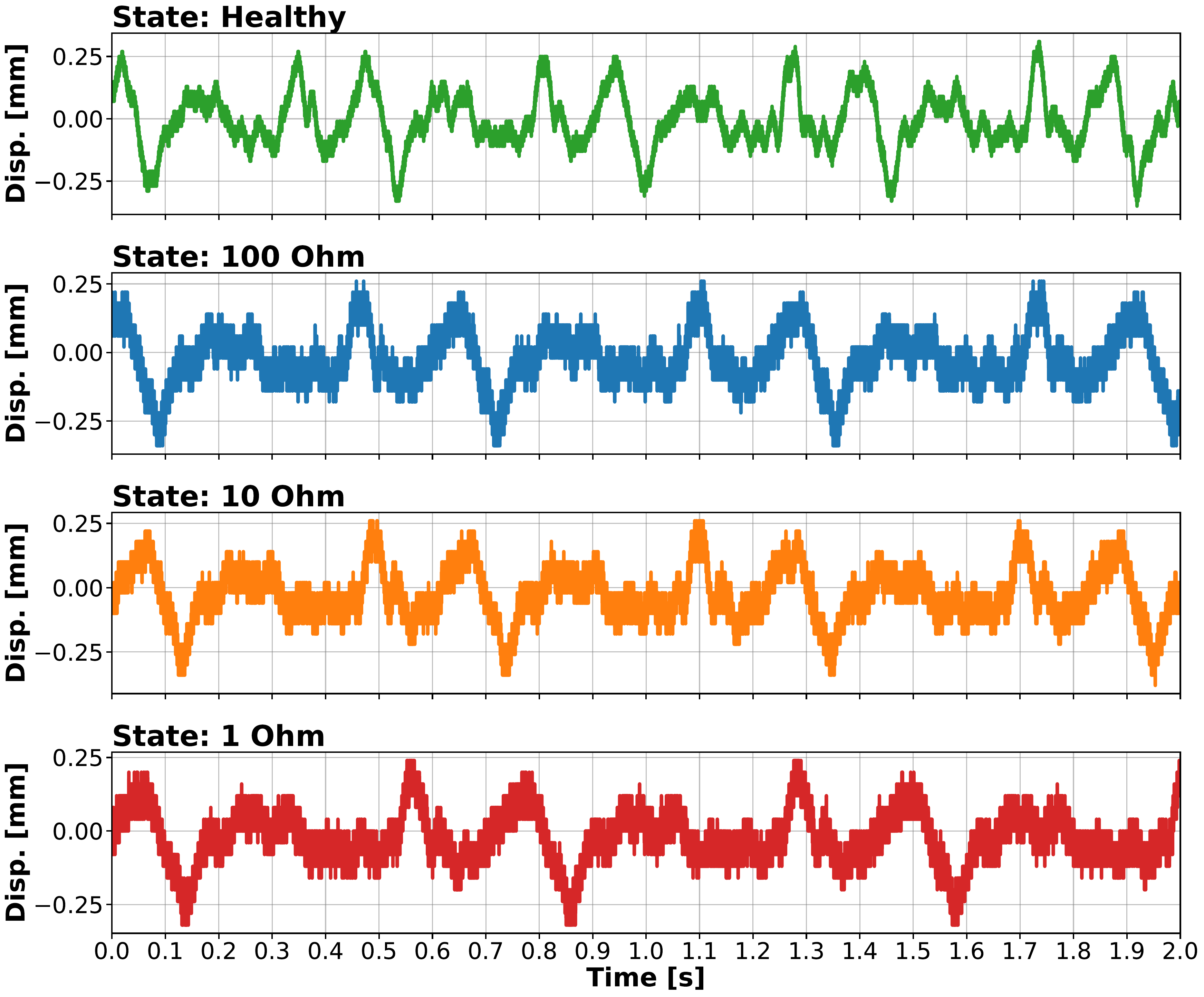}
    \caption{Representative filtered time-domain vibration displacement signals for the four investigated machine states.}
    \label{fig:time_domain2}
\end{figure}

A comparison of the acquired vibration signatures for aforementioned conditions is presented in Fig.~\ref{fig:time_domain2}. Despite varying fault severity, the peak-to-peak amplitude remains at a comparable level. This indicates that simple amplitude-based metrics are insufficient for reliable fault severity discrimination.

\subsection{Classification Performance}
\begin{figure}[!t]
    \centering
    \includegraphics[width=0.72\linewidth]{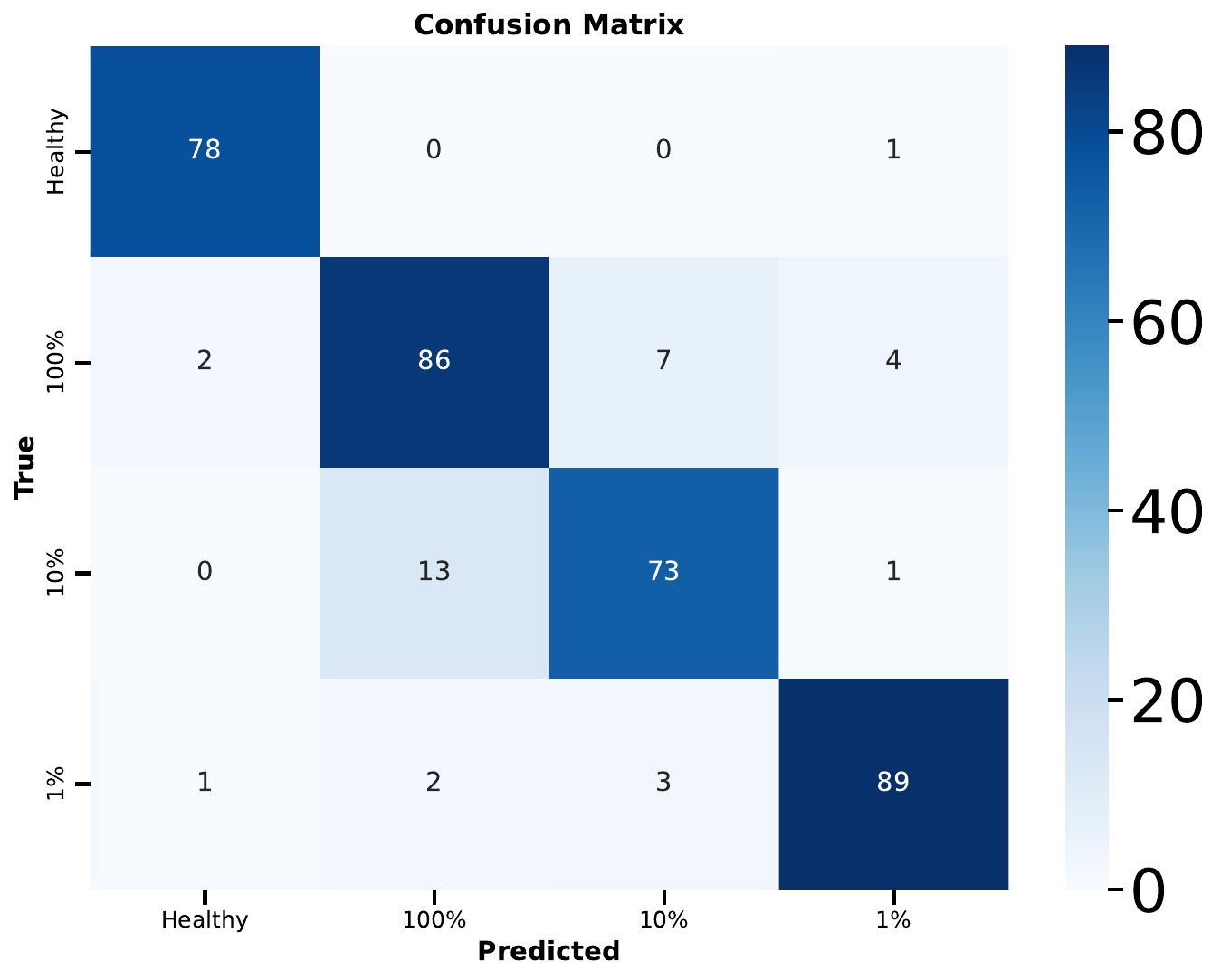}
    \caption{Confusion matrix showing 99\% healthy recall with balanced multi-class performance. Most misclassification occur between adjacent severity levels.}
    \label{fig:confusion_matrix2}
\end{figure}

An XGBoost classifier  \cite{ChenGuestrin2016} was trained using 
the 18-dimensional hybrid feature vector (8 spectral, 7 temporal/statistical, 2 current, 1 operating parameter). Leave-one-out cross-validation achieved 90.56\% overall accuracy on 360 measurements. The confusion matrix (Fig.~\ref{fig:confusion_matrix2}) and per-class metrics (Table~\ref{tab:classification_results2}) reveal excellent healthy-state detection with balanced performance across fault severities.

\begin{table}[htbp]
\centering
\caption{Per-Class Classification Performance (LOO Cross-Validation)}
\label{tab:classification_results2}
\begin{tabular}{lccc}
\hline
\textbf{Condition} & \textbf{Precision} & \textbf{Recall} & \textbf{F1-Score} \\
\hline
Healthy      & 0.96 & 0.99 & 0.97 \\
100 $\Omega$ & 0.85 & 0.87 & 0.86 \\
10 $\Omega$  & 0.88 & 0.84 & 0.86 \\
1 $\Omega$   & 0.94 & 0.94 & 0.94 \\
\hline
\textbf{Macro Avg.} & \textbf{0.91} & \textbf{0.91} & \textbf{0.91} \\
\hline
\end{tabular}
\end{table}
\subsection{Feature Importance Analysis}
Feature importance analysis (Table~\ref{tab:feature_importance2}) reveals balanced multi-domain contributions: current statistics (19.58\%), temporal complexity (19.16\%), operating parameters (6.07\%), and spectral/statistical features (55.19\%). The top-ranked features include current standard deviation (11.82\%), Hjorth Complexity (11.73\%), and mean current (8.76\%), demonstrating that no single domain dominates classification performance.
\begin{table}[htbp]
\centering
\caption{Top 10 Features by XGBoost Importance}
\label{tab:feature_importance2}
\begin{tabular}{lcc}
\hline
\textbf{Feature} & \textbf{Importance} & \textbf{Type} \\
\hline
Std Current         & 11.82\% & Current \\
Hjorth Complexity   & 11.73\% & Temporal \\
Mean Current        & 8.76\%  & Current \\
Perm. Entropy       & 7.43\%  & Temporal \\
Rotation Frequency  & 6.07\%  & Operating \\
RMS Total           & 5.37\%  & Spectral \\
RMS $2 \times f_r$  & 5.29\%  & Spectral \\
Peak-to-Peak        & 5.18\%  & Spectral \\
RMS $8 \times f_r$  & 4.99\%  & Spectral \\
Skewness            & 4.96\%  & Statistical \\
\hline
\end{tabular}
\end{table}

The multi-domain feature set suggests that current statistics may provide useful operational context (standard deviation captures load variations), while Hjorth Complexity appears to capture aspects of nonlinear temporal structure  not readily observable in frequency analysis. Spectral features across multiple harmonics (2×, 4×, 6×, 8× $f_r$) may further contribute to capturing distributed fault signatures.

The ±1.0\,Hz integration windows around mechanical harmonics effectively capture ITSC-induced spectral changes while maintaining robustness against rotation frequency estimation errors (typical $\Delta f_r < 0.1$\,Hz). This bandwidth encompasses the fundamental sideband structure predicted by UMP theory while avoiding excessive noise contamination from distant frequency bins.

For measurements at the dominant frequency range ($f_r \approx 1.5$--2.0\,Hz), the 1.0\,Hz window captures both the carrier energy and near-carrier modulation sidebands that characterize fault progression. The narrow sideband detection windows (±0.1\,Hz) specifically target amplitude modulation components at the rotation frequency, providing complementary fine-resolution diagnostic information.

\subsection{Hybrid Features and Diagnostic Performance}

The hybrid feature vector balances interpretability with discriminative power (Table~\ref{tab:feature_importance2}). Current statistics provide operational context, temporal complexity quantifies nonlinear signal structure, statistical features characterize morphology changes, and spectral features capture distributed fault signatures across harmonics. The achieved performance (Table~\ref{tab:classification_results2}) demonstrates  early-stage detection with high recall across all severities, enabling prognostic maintenance scheduling based on fault progression rather than binary presence/absence detection.

\subsection{Comparison with Existing Methods}

Compared to recent binary detection methods (e.g., Fang et al. \cite{FANG2023316}: 97\% accuracy using time-series prediction), our four-class severity classifier achieves 90.56\% overall accuracy while providing diagnostic granularity unavailable in binary frameworks. The framework maintains 99\% healthy recall (exceeding binary performance) while additionally distinguishing between three fault severity levels.

This multi-class capability supports actionable prognostic maintenance, as early-stage fault detection may enable planned generator outage. Furthermore, the hybrid feature design relies on physically meaningful spectral and temporal descriptors rather than end-to-end black-box time-series representation, which may facilitate  industrial adoption in applications where diagnostic transparency is important for operator trust and regulatory compliance.

\section{Conclusion}

This paper presented a multi-class ITSC severity classification framework for synchronous generators achieving 90.56\% overall accuracy on 360 measurements, with 99\% healthy recall and 87\% mild fault recall demonstrating effective early-stage detection using minimal sensor infrastructure.

Key contributions include: (i) hybrid 18-dimensional feature extraction  combining spectral, temporal complexity, statistical, and operational 
features with balanced multi-domain contributions, (ii) multi-class severity classification achieving 90.56\% accuracy with 99\% healthy 
recall and 87\% mild fault detection, (iii) 
a diagnostic framework based on physically meaningful features, promoting transparency in maintenance decisions, and (iv) practical 
deployment using minimal sensor infrastructure (single displacement sensor + current probe).

The results demonstrate that the proposed multi-domain diagnostic framework can achieve reliable ITSC severity discrimination under controlled laboratory conditions using standard industrial sensing hardware. The achieved performance indicates strong potential for practical condition-based maintenance applications in synchronous generators. However, further validation across different machine types, power rating, and real industrial environments would help assess robustness and generalization capability.


\section*{Acknowledgment}
This research has been supported by the European Union and the Ministry of Education, Youth and Sports of the Czech Republic under the project OP JAK Czech Incubator of Technologies for Energy Networks, number CZ.02.01.01/00/23\_020/0008490. This work was accepted and presented at the 2026 IEEE International Symposium on Industrial Electronics (IEEE ISIE) in Nagoya, Japan, in June 2026. It is included in the conference proceedings, and the copyright has been transferred to IEEE.

\bibliographystyle{IEEEtran}

\bibliography{ref}

@article{FANG2023316,
title = {Fault diagnosis of inter-turn short circuit in turbogenerator rotor windings based on vibration-current signal fusion},
journal = {Energy Reports},
volume = {9},
pages = {316-323},
year = {2023},
note = {2022 2nd International Joint Conference on Energy and Environmental Engineering},
issn = {2352-4847},
doi = {https://doi.org/10.1016/j.egyr.2023.03.019},
author = {Ruiming Fang and Zilin Liu and Changqing Peng and Yulei Yang and Shuming Zhang}
}

@ARTICLE{VirtualPower,
  author={Yucai, Wu and Yonggang, Li},
  journal={IEEE Transactions on Energy Conversion}, 
  title={Diagnosis of Rotor Winding Interturn Short-Circuit in Turbine Generators Using Virtual Power}, 
  year={2015},
  volume={30},
  number={1},
  pages={183-188},
  doi={10.1109/TEC.2014.2339300}}

@ARTICLE{Albright,
  author={Albright, D. R.},
  journal={IEEE Transactions on Power Apparatus and Systems}, 
  title={Interturn Short-Circuit Detector for Turbine-Generator Rotor Windings}, 
  year={1971},
  volume={PAS-90},
  number={2},
  pages={478-483},
  doi={10.1109/TPAS.1971.293048}}

@inproceedings{ShaftVoltage,
author = {Yucai, Wu and Yonggang, Li and Heming, Li},
year = {2012},
month = {10},
pages = {1-6},
title = {Diagnosis of turbine generator typical faults by shaft voltage},
isbn = {978-1-4673-0330-9},
journal = {Conference Record - IAS Annual Meeting (IEEE Industry Applications Society)},
doi = {10.1109/IAS.2012.6374011}
}

@article{Tavner,
author = {Tavner, P.J. and Ran, Li and Penman, Jim and Sedding, Howard},
year = {2008},
month = {01},
pages = {},
title = {Condition Monitoring of Rotating Electrical Machines},
isbn = {9780863417412},
journal = {Bibliovault OAI Repository, the University of Chicago Press},
doi = {10.1049/PBPO056E}
}

@INBOOK{Klempner,
  author={Klempner, Geoff and Kerszenbaum, Isidor},
  booktitle={Handbook of Large Turbo-Generator Operation and Maintenance}, 
  title={Monitoring and Diagnostics}, 
  year={2018},
  volume={},
  number={},
  publisher={IEEE},
  pages={331-428},
  doi={10.1002/9781119390718.ch5}}

@article{NandiReview2005,
  author  = {S. Nandi and H. A. Toliyat and X. Li},
  title   = {Condition Monitoring and Fault Diagnosis of Electrical Motors—A Review},
  journal = {IEEE Transactions on Energy Conversion},
  year    = {2005},
  volume  = {20},
  number  = {4},
  pages   = {719--729},
  doi     = {10.1109/TEC.2005.847955},
  url     = {https://doi.org/10.1109/TEC.2005.847955}
}

@inproceedings{ChenGuestrin2016,
  author       = {T. Chen and C. Guestrin},
  title        = {XGBoost: A Scalable Tree Boosting System},
  booktitle    = {Proceedings of the 22nd ACM SIGKDD International Conference on Knowledge Discovery and Data Mining},
  year         = {2016},
  pages        = {785--794},
  doi          = {10.1145/2939672.2939785},
  url          = {https://doi.org/10.1145/2939672.2939785}
}

@article{Hjorth1970,
    author = {Hjorth, Bo},
    title = {EEG Analysis Based on Time Domain Properties},
    journal = {Electroencephalography and Clinical Neurophysiology},
    year = {1970},
    volume = {29},
    number = {3},
    pages = {306--310},
    doi = {10.1016/0013-4694(70)90143-4}
}

@article{Bandt2002,
    author = {Bandt, Christoph and Pompe, Bernd},
    title = {Permutation Entropy: A Natural Complexity Measure for Time Series},
    journal = {Physical Review Letters},
    year = {2002},
    volume = {88},
    number = {17},
    pages = {174102},
    doi = {10.1103/PhysRevLett.88.174102}
}

@article{Rajabi2022,
    author = {Rajabi, S. and Azari, M.S. and Santini, S. and Flammini, F.},
    title = {Fault Diagnosis in Industrial Rotating Equipment Based on Permutation Entropy, Signal Processing and Multi-Output Neuro-Fuzzy Classifier},
    journal = {Expert Systems with Applications},
    year = {2022},
    volume = {206},
    pages = {117754},
    doi = {10.1016/j.eswa.2022.117754}
}

@article{HeEnergies2023,
  author  = {He, Yuling and Jiang, Mengya and Sun, Kai and Qiu, Minghao and Gerada, David},
  title   = {Analysis on Rotor Vibration Characteristics under Dynamic Rotor Interturn Short Circuit Fault in Synchronous Generators},
  journal = {Energies},
  year    = {2023},
  volume  = {16},
  number  = {18},
  pages   = {6585},
  doi     = {10.3390/en16186585}
}

@article{FangEnergyReports2023,
  author  = {Fang, Ruiming and Liu, Zilin and Peng, Changqing and Yang, Yulei and Zhang, Shuming},
  title   = {Fault diagnosis of inter-turn short circuit in turbogenerator rotor windings based on vibration-current signal fusion},
  journal = {Energy Reports},
  year    = {2023},
  volume  = {9},
  number  = {Supplement 2},
  pages   = {316--323},
  doi     = {10.1016/j.egyr.2023.03.019}
}

@inproceedings{dosSantosCBMAG2022,
  author    = {dos Santos, Guilherme Felipe and Wengerkievicz, Carlos Alexandre Corr{\^e}a and Batistela, Nelson Jhoe and Bernard, Laurent Didier and de Freitas, Luciano Mendes and Borges, Luiz Ant{\^o}nio Campos and Matsuo, Tiago Kaoru and Tominaga, Luan},
  title     = {Detection of Electrical Faults in a Synchronized Generator through Mechanical Vibration at Various Operating Conditions},
  booktitle = {Anais do 15{\textordfeminine} Congresso Brasileiro de Eletromagnetismo (CBMAG)},
  year      = {2022},
  publisher = {Even3},
  note      = {Online proceedings (ISBN: 978-65-5941-681-3)}
}

@article{YuanMPE2021,
  author  = {Yuan, Xing-Hua and He, Yu-Ling and Liu, Manqiang and Wang, Hui and Wan, Shi-Tong and Vakil, Gaurang},
  title   = {Impact of the Field Winding Interturn Short-Circuit Position on Rotor Vibration Properties in Synchronous Generators},
  journal = {Mathematical Problems in Engineering},
  year    = {2021},
  volume  = {2021},
  pages   = {9236726},
  doi     = {10.1155/2021/9236726}
}
\end{document}